\documentstyle[prl,twocolumn,aps]{revtex}
\input{epsf}
\begin{document}
\draft
\twocolumn[\hsize\textwidth\columnwidth\hsize\csname @twocolumnfalse\endcsname
\title{Critical Spectral Statistics at the Metal-Insulator Transition in Interacting Fermionic Systems}

\author{Philippe Jacquod}

\address {Department of Applied Physics, P. O. Box 208284, Yale University, New Haven, CT 06520-8284}

%\date{\today}
\date{june 29, 1998}

\maketitle

\begin{abstract}
The spectral properties of a disordered system with few interacting 
three-dimensional spinless fermions are investigated. We show the 
existence of a critical spacings distribution which is invariant upon 
increase of the system size, but strongly depends on the number of 
particles. At the critical point, we report a substantial decrease 
of the degree of level repulsion as the number of particles increases 
indicating a decrease of nearest level correlations associated with the 
sparsity of the Hamiltonian matrix.
\end{abstract}
\vskip1pc]

%\begin{multicols}{2}
\narrowtext

% Introduction
%insert
The presence and strength of a disorder potential strongly modify
the influence of electronic interactions on localization. 
While in clean systems repulsive interactions always
localize, their effect can be reversed by the presence of a strong
disorder. There, interaction-induced hoppings between localized states 
may help electrons to overcome the random potential thus reducing 
localization \cite{efros,joe,tala}. In this paper we will show that repulsive 
interactions can indeed trigger a metal-insulator transition in a few-particle
strongly disordered system, thereby turning a one-body insulator into a 
many-body metal. 
%insert

The statistical properties of spectra of disordered one-particle 
systems are 
known to be closely related to the localization properties of the 
corresponding eigenstates \cite{altar,shklo,kravtsov}. 
In the localized phase,
states which are close in energy have an exponentially small overlap. 
Consequently the levels $E_i$ are uncorrelated and the corresponding
normalized  
spacings $s_i \equiv (E_{i+1}-E_i)/\langle E_{i+1}-E_i \rangle$ are
distributed according to the Poisson distribution $P_P(s) = \exp(-s)$. 
On the other hand, the large overlap of delocalized eigenstates of neighboring 
energy induces correlations in the spectrum and leads to level repulsion, 
%insert
so that in the time-reversal symmetric case, 
%insert
the distribution of energy spacings corresponds
to the Wigner-Dyson distribution $P_{WD}(s) = \frac{\pi}{2} s 
\exp(-\pi s^2/4)$ \cite{altar}. While these two limiting behaviors set in 
progressively as the linear system size $L$ increases, 
%insert
%insert
a third, size-independent distribution 
%insert
%insert
appears at the metal-insulator transition due to the divergence of the 
correlation length \cite{shklo} . This behavior suggested 
a new powerful method to locate the metal-insulator transition in one-particle
models, 
%insert
relying only on the knowledge of the spectrum
%insert
to deduce the localization properties of the eigenfunctions.
It is one of the purposes of the present letter to generalize this method to 
many-body interacting disordered systems.

Spectral properties of many-body systems have already attracted some interest.
Early works showed 
%insert
%insert
that, for clean systems, only non-generic models where 
special group symmetries ensure the integrability, have a Poissonian 
level spacings distribution \cite{bel}. 
In the presence of disorder, it has been
suggested that the interaction induces
correlations between many-body wave functions and in the corresponding 
spectrum, thus resulting in 
level repulsion. This phenomenon has been related to a
delocalization in real space in the Coulomb glass \cite{tala} and in
a system of two interacting particles in a disordered potential \cite{wpich},
and to the recently intensively investigated {\it delocalization
in Hilbert/Fock space} \cite{pich,m2body,m2body2}.
The latter concept refers to an interaction-induced spreading of the 
many-body states over the basis of Slater determinants built out of the 
one-particle states. For sufficiently large interactions, this spreading
covers many basis states whose number represents a localization length in
Hilbert/Fock space much in the usual sense. 
The local spectral density typically acquires a Breit-Wigner shape
$ \rho_{BW}(E) = \Gamma/( 2 \pi(E^2+\Gamma^2/4))$
and the spreading of many-body states may be characterized by 
the width $\Gamma$ \cite{wpich,jac}. The latter may be estimated from the
Golden Rule which gives $\Gamma \sim Q^2 /\Delta_{c}$ where
$ Q^2 $ is the mean square interaction matrix element and
$\Delta_{c}$ the spacings between states directly coupled by the two-body
interaction. Thus the 
interaction starts to mix directly coupled states as the critical threshold 
$Q_c \sim \Delta_{c}$ is reached 
and simultaneously, spectral correlations
result in level repulsion and the emergence of quantum chaos
\cite{m2body}. Quite surprisingly, it was shown in
%insert
%insert 
\cite{m2body2} that the
number of basis states occupied by a single interacting
state is given by $\xi \sim \Gamma/\Delta_{n}$ where $\Delta_n$ is the
$n$-particle mean spacing. As quantum
chaos starts to set in one has
$\xi_c \sim \Delta_c/\Delta_n \gg 1$
so that contrarily to the common belief, level repulsion appears {\it after}
many noninteracting states have been mixed by the interaction. 
Since the models studied in \cite{m2body,m2body2} contained no information
on the real space, no conclusion could be drawn as to a transition
in real space associated to the emergence of quantum chaos or the mixing of
eigenstates, a question which is of particular interest.

%insert
In this letter, we present a systematic investigation of the spectral
properties of a system of few interacting quasiparticles
above a frozen (i.e. noninteracting) Fermi sea. 
The neglect of interaction processes involving one-particle states
located in the frozen Fermi sea is advantageous in two respects : 
(i) it allows to
reduce significantly the Hilbert space volume and consequently to 
study large systems of linear sizes up to $L=10$ in three dimensions 
(Presumably,
it is the impossibility of studying systems with large enough sizes that
hindered investigations similar to \cite{shklo} in interacting systems 
so far.) and
(ii) it projects out one-body tail states which are known to be strongly
localized and hence much less sensitive to the interaction-induced 
delocalizing effect. 
For the two-quasiparticle case, 
this model was originally proposed in \cite{joe} and further studied in 
\cite{tiq}. 

More specifically, we consider
spinless fermions in a disordered Anderson-like model
with a repulsive nearest neighbor interaction of 
strength $U$ whose Hamiltonian is
\begin{eqnarray}
H & = & P_b \left[\sum_i W_i 
a^\dagger_i a_i + \sum_{\langle i;j \rangle} (U a^\dagger_i a^\dagger_j 
a_j a_i -t a^\dagger_i a_j) \right] P_b
\end{eqnarray}
Here, $P_b$ projects out Slater determinants containing
one-particle states of energy lower than the threshold energy $E_b$. 
$\langle i;j \rangle$ restricts the sum to nearest-neighbor sites on a 
three-dimensional cubic lattice and
$ W_i$ is the on-site disorder randomly distributed between $-W/2$ and $W/2$.
In our numerical investigations we fixed $E_b \approx -3.3 t$ 
corresponding to a filling factor of $\nu =1/3$ in the disorder regime
we will consider $W=18$. 
We point out that this choice is arbitrary and checked numerically that
the physical picture presented here does not depend on this particular choice 
of $E_b$.
%insert
For a system of linear size $L$ and $n$ quasiparticles, the ground-state 
energy is approximately
$\epsilon_g \approx n E_b + n(n+1) \Delta_1/2 + 
3 U n(n-1) \Delta_1/ B_1$.
$\Delta_1 \approx B_1/L^d$ is
the one-particle average spacings and $B_1 \approx 2dt + W$ is
the one-particle bandwidth. We also
define the many-body
excitation energy as $\delta \epsilon \equiv \epsilon-\epsilon_g$. 
As the introduction of the frozen Fermi sea 
may appear somehow artificial, 
we stress that for low many-body energy excitations 
$\delta \epsilon$ and moderate interaction 
$U \ll B_1$, only
particles in an energy layer of size $T \sim \sqrt{\delta \epsilon\Delta_1}$ 
around the Fermi
level effectively interact and their number is given 
by $n_{eff} \sim T/\Delta_1 \sim  \sqrt{\delta \epsilon/\Delta_1}$ 
so that for $n_{eff} < n$ our model is fully justified. 

The numerical investigations were carried out much in the same way as in
reference \cite{shklo,m2body}. We computed the distribution 
$P(s)$ for the many-body level spacings $s_i \equiv (\epsilon_{i+1}-\epsilon_i)/\Delta_n$
for $n=2-6$ quasiparticles, at fixed excitation energies and disorder, 
for different interaction strengths and
linear system size. As the interaction strength increases,
more and more states are coupled and we expect quantum chaos to set in,
i.e. the level
spacings distribution undergoes a crossover from Poisson to Wigner-Dyson
distribution. To quantitatively study this crossover we computed
the value of the parameter $\eta = \int_0^{s_0}[P(s)-P_{WD}(s)]ds/ \int_0^{s_0}[P_P(s)-P_{WD}(s)]ds$. Here, $s_0 \approx 0.473$ is the smallest root of
$P_P(s) = P_{WD}(s)$ and $\eta$ varies between 1 
($P(s) = P_P(s)$) and 0 ($P(s) = P_{WD}(s)$). All investigations were carried
out at fixed disorder $W=18.$ slightly above the critical disorder 
$W_c \approx 16.5$ of the Anderson model with boxed distributed disorder
\cite{shklo,zhar}. At this disorder, one-particle states still have a large
enough localization length $l_1$ so that the short-range interaction couples
a large number of states. 
We averaged over up to 10000 disorder realisations, so as
to get at least $N = 50000$ levels in an energy interval 
$[\delta \epsilon-\Delta_{\epsilon}/2;\delta \epsilon+\Delta_{\epsilon}/2]$, 
$\Delta_{\epsilon} < \Delta_1$.

\begin{figure}
\epsfxsize=3.3in
\epsfysize=2.5in
\epsffile{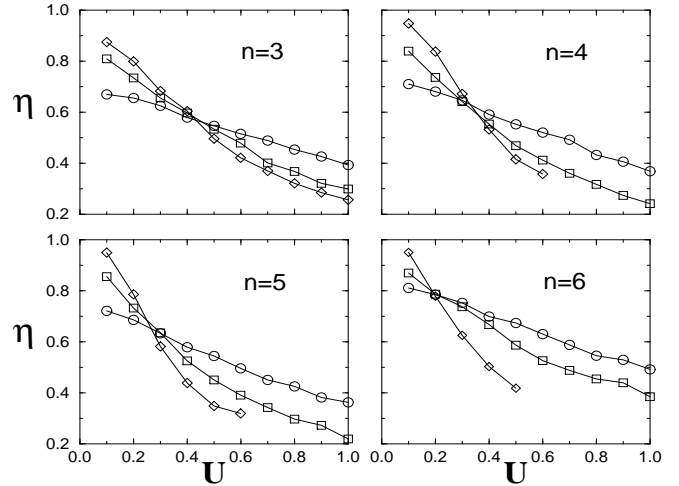}
\vglue 0.1cm
%\medskip
\caption{Evolution of the crossover parameter $\eta$ for $n=3$, 
$\delta \epsilon = 0.35$, $n=4$, $\delta \epsilon = 0.5$, 
$n=5$, $\delta \epsilon = 0.65$ and $n=6$, $\delta \epsilon = 0.8$ 
and disorder $W=18.$ 
(From left to right and top to bottom). The linear system sizes 
are $L=6$ (circles), $L=8$ (squares) and $L=10$ (diamonds). For each $n$,
excitation energies satisfy $n_{eff} \approx 
2 \sqrt{\delta \epsilon/\Delta_1} \approx n$ for $L=6$.}
\label{fig1}
\end{figure}

Fig.1 shows the 
dependence of the crossover parameter $\eta$ as a function of $U$ for 
different number of particles. Clearly, the situation is similar to the one
encountered in \cite{shklo} for the Anderson transition :
the $\eta$-curves belonging to different system sizes intersect at a critical
$U_c$ where $\eta_c$ is size-independent. 
This indicates the existence of a scaling due 
to the divergence of a correlation length associated to the many-body 
eigenstates: there is a continuous 
transition from an insulating phase at $U<U_c$ where $\eta$ increases with
the system size to a metallic one
at $U>U_c$ where the opposite behavior is observed. 
%insert
We stress that, as already observed in different strongly disordered 
systems \cite{efros,joe} and unlike in clean ones, 
the presence of a repulsive interaction {\it favors} delocalization. 
%insert

The existence of this transition dictates the
value of $\eta_c$ which consequently depends on the
number of particles. Thus the critical distribution, i.e. the degree
of level repulsion also depends on the number of particles,
as shown in Fig.2. 
$\eta_c(n)$ is a monotonously increasing
function of $n$ and it is tempting to say that as $n \rightarrow \infty$,
the critical distribution tends to $P_P(s)$, i.e. {\it there is no level
repulsion at the metal-insulator transition in a system with an infinite
number of interacting quasiparticles}.

The increase of $\eta_c(n)$ with $n$ and the corresponding 
decrease in the degree of correlations between nearest levels may indicate
that this metal-insulator transition proceeds much in the same way
as the transition to chaos reported in \cite{m2body}, i.e. that many-body
delocalization results from the mixing of directly
coupled states and that transport in the metallic phase proceeds
by interaction-induced jumps between these
states. In this case, the critical interaction satisfies the condition
$ Q_c \sim \Delta_c $.
Accordingly, only states separated by an energy interval 
$\Delta_c \gg \Delta_n$ are correlated and the resulting 
level repulsion can be distributed over $N_n \equiv \Delta_c/\Delta_n$
successive level spacings. Hence if $N_n \rightarrow \infty$, 
as $n \rightarrow \infty$, the level repulsion vanishes at the transition.

\begin{figure}
\epsfxsize=3.25in
\epsfysize=2.6in
\epsffile{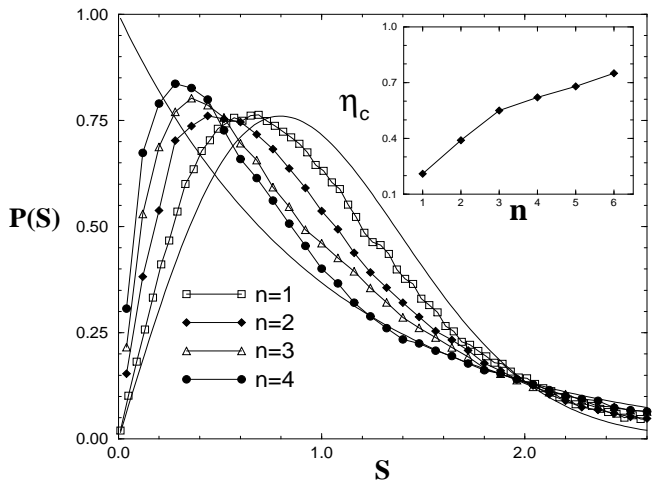}
\vglue 0.2cm
%\medskip
\caption{Critical distributions for $n=1$ (squares), $n=2$ (full diamonds),
$n=3$ (triangles) and $n=4$ (full circles). 
Also shown are the Wigner-Dyson and the Poisson distribution (full lines).
Inset : Evolution of the critical crossover parameter $\eta_c$ as a function
of the number of particles.} 
\label{fig2}
\end{figure}

We can estimate a lower boundary for 
$N_n$ following a similar argument as in \cite{m2body}. 
Assuming that at the transition, the system is
thermalized, we consider $n \rightarrow \infty$ particles at a temperature 
$1 \ll T/\Delta_1 \ll n $. The number of effectively interacting particles
is then $n_{eff} \sim T/\Delta_1$ so that they may be distributed 
over $m \approx 2 n_{eff}$ different one-body levels. 
The two-body interaction connects each state to roughly
$K \approx n_{eff}(n_{eff}-1)(m-n_{eff})(m-n_{eff}-1)/4 \sim n_{eff}^3
(T/\Delta_1)$ states. 
All these possible transitions occur in an energy interval
$B_2 \approx (2m-4) \Delta_1 \approx 4 n_{eff} \Delta_1$
so that in agreement with \cite{m2body}
\begin{equation} 
\Delta_c \sim \frac{B_2}{K} \sim \frac{\Delta_1^2}{n_{eff}^2 T}
\end{equation}
Next we have 
$\Delta_n \approx 2 \Delta_1^n (n-1)!/\delta 
\epsilon^{n-1}$ (as before $\delta \epsilon \approx T^2/\Delta_1$)
and consequently $N_n$ is exponentially large 
\begin{eqnarray}
\ln (N_n) & \approx & n_{eff} \ln(n_{eff})
\end{eqnarray}
This latter result means in particular that at the transition, level
correlations are distributed over a fast increasing number of many-body levels,
and thus explains how the level repulsion is reduced by an increasing number
of interacting particles. Equation (3) follows from the
increasing sparsity of the Hamiltonian matrix with increasing number of 
particles. 

Better estimates of $N_n$ and $\Delta_c$ 
should take into account the fact that, 
close to the Anderson transition, the structure of the 
one-particle eigenstates reduces the number of effectively
coupled states, while enhancing their correlations and hence 
the mean transition matrix elements \cite{blanter}. The latter in their turn,
are directly related to the spacing between directly coupled states by
the relation obeyed at the transition $Q_c \sim \Delta_c$. 

\begin{figure}
\epsfxsize=3.25in
\epsfysize=2.6in
\epsffile{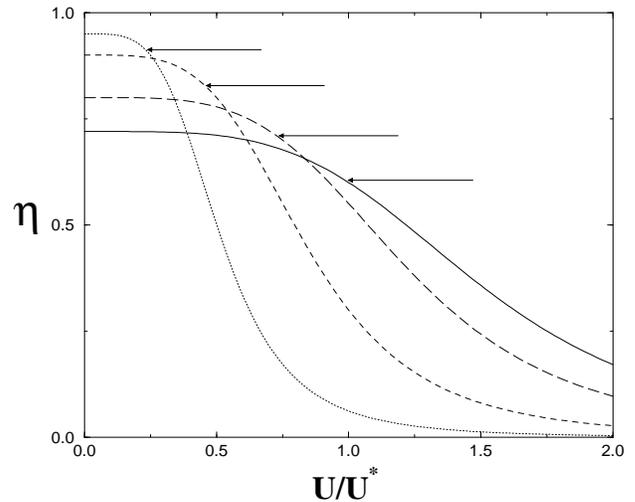}
\vglue 0.3cm
%\medskip
\caption{Schematic evolution of $\eta$ as a function
of the renormalized interaction $U/U^*$ in the regime 
$n_{eff} \approx 2 \sqrt{\delta \epsilon/\Delta_1} <n$
($U^*$ corresponds to the smaller system size.).
The effective number of interacting particles increases with the system size,
and in the process, $\eta_c(n_{eff})$ increases as indicated by the 
arrows and the
curves no longer intersect at one point. In the other regime studied 
in this article,
$\delta \epsilon > \Delta_1 n$, the number of effectively interacting particle
is invariant upon increase of the system size and the transition is
unambiguously determined by the intersection of all $\eta$-curves 
(see Fig.1).}
\label{fig3}
\end{figure} 

Our interpretation relates the level repulsion at the transition to the
$n$-body correlation volume $B_1/\Delta_c$, i.e. the volume inside which
$n$-body states have a strong overlap and are therefore directly connected
by the interaction. 
Indeed in the three-dimensional 
one particle Anderson model, this volume goes like $L^{d-\omega}$ 
with a multifractal exponent $\omega \approx 1.3$ \cite{brandes}. 
Hence we have $N_1 \sim L^{\omega}$ and
therefore the critical statistics is intermediate between Poisson 
($N_1 \approx L^d$) and Wigner-Dyson ($N_1 \approx 1$).
Moreover in the case of two particles, the two-body interaction connects all
states together and $N_2$ can be reduced only by the
one-particle correlations. Consequently,
the critical statistics is very well fitted by the
semi-Poisson $P_{SP}(s) = 4 s \exp(-2s)$, as can be 
the case for the one-particle model \cite{remark}. As $n$ increases,
the Hamiltonian matrix becomes sparser and sparser and the critical
spacings distribution exhibits less and less level repulsion.

In the thermodynamic (TD) limit, for such Fermi systems, 
it is natural to define a temperature
as $n_{eff} T = T^2/\Delta_1 \sim \delta \epsilon$, 
and as $T$ must be constant as $L \rightarrow \infty$, 
this implies that 
$T/B_1 \approx n_{eff} \Delta_1/B_1= {\rm const}$.
Two different regimes are then of interest: \\
$\bullet$ The one-particle critical regime close to the Anderson transition, 
where the system size is significantly smaller than the one-particle 
localization length. There, both
the one-particle correlation volume and
$n_{eff}$ increase as $\sim L^{d-\omega}$ until $L \approx l_1$ and
since $l_1$ can be very large, so is $n_{eff}$ and
there is almost no level repulsion at the transition in the TD limit. 
In this case the scaling at the transition is not
observable since $\eta_c$ increases with $n$ 
and the $\eta$-curves intersect at different points (see Fig.3). \\
$\bullet$ The one-particle localized regime $1 \ll l_1 < L$, where
we have $\Delta_1 \approx B_1/l_1^d = {\rm const}$.
This regime corresponds to the results presented in Fig.1 for which $n_{eff}$
is constant and one has a fixed-point at a well-defined value of $U_c$.

Since it is expected that a large interaction will ultimately transform the
system back into an insulator, it is important to estimate the parametrical
width of the metallic phase. In this case, it seems natural to carry out
the substitution $\Delta_1 \rightarrow Q$ in (2) so that in the localized
regime with $n_{eff} = {\rm const}$, the condition 
$Q_c \sim \Delta_c$ leads to a metallic phase in the range
\begin{equation}
\frac{\Delta_1}{n_{eff}^2 T} \leq \frac{Q}{\Delta_1}
\leq \frac{n_{eff}^2 T}{\Delta_1}
\end{equation}
Since $n_{eff}^2 T = \delta \epsilon \gg \Delta_1$, 
the regime defined by (4) is
parametrically large enough to allow for a metallic phase.
The backtransition at high $Q$ occurs before the 
Wigner crystallization, i.e. it is induced by the lowering of the density of
states much in the same way as the transition described above, 
and occurs before the opening of a gap.

Without the constraint of a filled Fermi sea ($P_b = 1$), 
levels are filled from the bottom 
of the Anderson band where the decrease of $l_1$ implies an increase of 
$\Delta_{1}$.
%insert
Correspondingly, the metallic regime determined by (4) is reduced unless
the disorder is lowered below
$W_c$ so as to get the one-particle mobility edge close to the Fermi level,
and increase the localization length of the one-particle band-tail states.
In this case, the situation should be similar to the one reported above,
i.e. the one-particle correlation volume is sufficiently large to
allow the mixing of a large number of states and a metal-insulator transition
should occur.
%insert

In conclusion we successfully applied the method developed 
in \cite{shklo} to disordered many-body systems. 
We showed the existence of a continuous 
metal-insulator phase transition where
the critical level spacings distribution is different from the
one at the Anderson transition.
These considerations quite naturally lead to the
identification of the metallic phase with the appearance of quantum chaos and
the mixing of directly coupled states.

It is a pleasure to thank A. D. Stone for many suggestions that 
helped improve this paper. Very interesting discussions with M. Janssen 
are also acknowledged.
Numerical computations were performed at the Swiss Center
for Scientific Computing. Work supported by the 
Swiss National Science Foundation.

%\end{multicols}


\begin{thebibliography}{99}
\bibitem{efros} D. L. Shepelyansky, Phys. Rev. Lett. {\bf 73}, 2607, (1994);
R. Berkovits and Y. Avishai, Europhys. Lett. {\bf 29}, 475,
(1995); E. V. Tsiper and A. L. Efros, Phys. Rev. B {\bf 57}, 6949, (1998).
\bibitem{joe}  Y. Imry, Europhys. Lett. {\bf 30}, 405 (1995).
\bibitem{tala} J. Talamantes, M. Pollak and L. Elam, Europhys. Lett. {\bf 35}, 511, (1996).
\bibitem{altar} B. L. Altshuler and B. I. Shklovskii, Sov. Phys. JETP {\bf 64}, 127, (1986).
\bibitem{shklo} B. I. Shklovskii, B. Shapiro, B. R. Sears, P. Lambrianides and
H. B. Shore, Phys. Rev. B {\bf 47}, 11487 (1993). 
\bibitem{kravtsov} V. E. Kravtsov, I. V. Lerner, B. L. Altshuler and A. G. Aronov, Phys. Rev. Lett. {\bf 72}, 888, (1994).
\bibitem{bel} G. Montambaux, D. Poilblanc, J. Bellissard and C. Sire, Phys.
Rev. Lett. {\bf 70}, 497 (1993); D. Poilblanc, T. Ziman, J. Bellissard, F.
Mila and G. Montambaux, Europhys. Lett. {\bf 22}, 537 (1993).
\bibitem{wpich} D. Weinmann and J.-L. Pichard, Phys. Rev. Lett. {\bf 77}, 1556, (1996). 
\bibitem{pich} B. L. Altshuler, Yu. Gefen, A. Kamenev and S. L. Levitov, 
Phys. Rev. Lett. {\bf 78}, 2803 (1997);D. Weinmann, J.-L. Pichard and Y. Imry, J. Phys. I France {\bf 7}, 1559, (1997); R. Berkovits and Y. Avishai, Phys. Rev. Lett. {\bf 80}, 568, (1998); J. Talamantes and A. M\"obius, 
Phys. Stat. Sol. {\bf 205}, 45, (1998).
\bibitem{m2body} Ph. Jacquod and D. L. Shepelyansky, Phys. Rev. Lett. {\bf 79}, 1837, (1997). 
\bibitem{m2body2} B. Georgeot and D. L. Shepelyansky, Phys. Rev. Lett. {\bf 79}, 4365, (1997).
\bibitem{jac} Ph. Jacquod and D. L. Shepelyansky, Phys. Rev. Lett. {\bf 75},
3501 (1995); Y. V. Fyodorov and A. D. Mirlin, Phys. Rev. B {\bf 52}, 11580, (1995); K. Frahm and A. M\"uller-Groeling, Europhys. Lett. {\bf 32}, 385, (1995).
\bibitem{tiq} F. von Oppen and T. Wettig, Europhys. Lett. {\bf 32}, 741, 
(1995); Ph.Jacquod and D.L.Shepelyansky, Phys. Rev. Lett. {\bf 78}, 4986, 
(1997); Ph. Jacquod, Phys. Stat. Sol. {\bf 205}, 263, (1998).
\bibitem{zhar} I. Zharekeshev and B. Kramer, Phys. Rev. B {\bf 52}, 13903, 
(1995).
\bibitem{remark} D. Braun, G. Montambaux and M. Pascaud, cond-mat/9712256. 
\bibitem{blanter} Ya. M. Blanter, Phys. Rev. B {\bf 54}, 12807 (1996);
A. D.Mirlin and Y. V.Fyodorov, Phys. Rev. B {\bf 55}, R16001, (1997); 
Ya. M. Blanter and A. D. Mirlin, Phys. Rev. E {\bf 55}, 6514, (1997).
\bibitem{brandes} T. Brandes, B. Huckenstein and L. Schweitzer, Ann. Phys. 
(Leipzig) {\bf 5}, 633, (1993).
%%%%%%%%%%%%%%%%%%%%%%%%%%%%%%%%%%%%%%%%%%%%%%%%%%%%%%%%%%%%%%%%%%%%%%%%%%%%%%%
\end{thebibliography}
\end{document}